\documentclass[12pt]{revtex4}
\usepackage{hyperref,amsmath}

\newcommand{\dg}{^\dagger}

\newcommand{\be}{\begin{equation}}
\newcommand{\ee}{\end{equation}}
\newcommand{\bes}[2]{\begin{equation}\begin{split}#1\end{split}\label{#2}\end{equation}}

\usepackage{tikz}
\usetikzlibrary{calc}
\usetikzlibrary{arrows}
\usetikzlibrary{decorations.markings}
\usepackage{pgfplots}
\usepackage{rotating}
\usetikzlibrary{pgfplots.groupplots}
\tikzset{
    axis break gap/.initial=0mm
}

\begin{document}

\title{Simulating Infinite Vortex Lattices in Superfluids}

\date{\today}
\author{Luca Mingarelli,  Eric E Keaveny,  and Ryan Barnett}
\affiliation{Department of Mathematics, Imperial College London,
London SW7 2AZ, United Kingdom}
\begin{abstract}
We present an efficient framework to numerically treat infinite periodic vortex lattices 
in rotating superfluids
described
by the Gross-Pitaevskii theory.  The commonly used split-step Fourier (SSF) spectral  methods
are inapplicable to such systems as the standard Fourier transform does not respect the 
boundary conditions mandated by the magnetic translation group.  We present a generalisation
of the SSF method which incorporates the correct boundary conditions 
by employing the so-called magnetic Fourier transform.  We test the method and show that it reduces
to known results in the lowest-Landau-level regime.  While we focus on rotating scalar superfluids 
for simplicity, the framework can be naturally extended to 
treat multicomponent systems and systems under more general `synthetic' gauge fields.

\end{abstract}
\maketitle

\section{Introduction}

Superfluid phases of matter have attracted interest over
many years and can occur in diverse physical systems including liquid
helium, neutron stars, and ultracold atomic gases \cite{babaev15}.
The response of a superfluid to mechanical rotation demonstrates one
of the most remarkable features of these systems.  Rather than rotate
like a classical fluid, a superfluid will instead nucleate quantised
vortices \cite{Onsager49, Feynman55} which carry angular momentum.  In
the limit of many vortices, a vortex lattice will form.  There exist
several thorough review articles and books addressing vortices in
superfluids including Refs.~\cite{Donnelly,Sonin87,Fetter09,Cooper08}.

Bose-Einstein Condensates (BECs) of ultracold atomic gases represent a
quintessential example of a superfluid.  Within Gross-Pitaevskii mean
field theory, these systems are described by a macroscopic wave
function $\psi$ and the system's energy in a rotating frame of
reference with a harmonic trapping potential is given by
\begin{align}
\label{energyfunctional}
E\left[\psi\right]
&=\int\text{d}x\text{d}y\left[\frac{\hbar^2}{2m}\left|\nabla\psi\right|^2+\frac{1}{2}m
  \omega^2 r^2|\psi|^2+\frac{g}{2}|\psi|^4
-\psi\dg\Omega L_z\psi-\mu |\psi|^2\right] \\
  &=\int\text{d}x\text{d}y\left[\frac{1}{2m}\left|\left(-i\hbar\nabla-{
          \bf A}_s\right)\psi\right|^2+\frac{1}{2} m \omega_{\rm
      eff}^2 r^2|\psi|^2+\frac{g}{2}|\psi|^4-\mu |\psi|^2\right].
\notag
\end{align}
In this equation, $m$ is the mass of the constituent bosons, $\omega$
is the frequency of the harmonic trapping potential,
$L_z=-i\hbar\left(x\partial_y-y\partial_x\right)$ is the angular
momentum operator, $\Omega$ is the rotational frequency, $g$ is the
interaction parameter, and $\mu$ is the chemical potential.  In the
second line of (\ref{energyfunctional}), the energy functional is
rearranged in a convenient way, where the gauge field ${\bf
A}_s=\Omega m(-y,x)$ and effective harmonic oscillator frequency
$\omega_{\rm eff}=\sqrt{\omega^2-\Omega^2}$ are introduced.  The
dynamics of the condensate is given by the corresponding
Gross-Pitaevskii Equation (GPE) $i \hbar \partial_t \psi = \delta E / \delta
\psi^*$.  We will focus on cases where the minimum of
(\ref{energyfunctional}) is a state with many vortices.  While there
is little doubt that such ground states correspond to triangular
vortex lattices, we note that multi-component systems, including
mixtures, spinor condensates \cite{Stamper13}, and multicomponent systems under
more general gauge fields \cite{Goldman14}
can have much richer vortex lattices.

In order to understand vortex solutions of the GPE, it is helpful to consider a number of length scales
naturally emerging from (\ref{energyfunctional}).  The healing length
defined as
$
\xi=\sqrt{\frac{\hbar^2}{2mg\bar{\rho}}}
$,
where $\bar{\rho}$ is the average superfluid density,
gives the characteristic core size of the vortices.  According to the
Feynman relation \cite{Feynman55}, the density of the vortices is
given by $\rho_v = \frac{m\Omega}{\pi \hbar}=\frac{1}{2 \pi
\ell_B^2}$ where, in analogy with quantum Hall systems, we have
introduced the `magnetic length'
$
\ell_B=\sqrt{\frac{\hbar}{2  m \Omega }}.
$
This length scale therefore provides the characteristic separation
between vortices.  Finally, $\omega_{\rm eff}$ sets the effective
trapping length scale
$
\ell_H= \sqrt{\frac{\hbar}{2  m \omega_{\rm eff} }}.
$
For weak interactions such that $\ell_H \ll \xi$, $\ell_{H}$ gives the
spatial extent of the condensate.  On the other hand, for stronger
interactions where $\ell_H \gg \xi$, the size of the condensate is
given by the Thomas-Fermi radius $R_{\rm TF} \sim \ell_H^2/\xi$.

Minimisation of (\ref{energyfunctional}) and its multi-component
extensions for the purpose of investigating vortex lattices typically
involves one of two approaches.  The first approach involves projecting the wave function $\psi$ into
the the lowest Landau level (LLL) and thereby writing 
$\psi(x,y) = \sum_n A_n \phi_n(x,y)$ where $\phi_n(x,y) \propto (x+iy)^n
e^{-(x^2+y^2)/(2\ell_B)^2}$ are lowest-Landau-level eigenfunctions.
The projected wave function is then inserted into (\ref{energyfunctional}) and
the energy is minimised with respect to the $A_n$-parameters
\cite{Cooper08}.  This method is appropriate for cases of finite as
well as zero effective trapping potential $\omega_{\rm eff}$, where
the latter case corresponds to a condensate of infinite spatial
extent.  Typically, this is the method of choice when investigating
infinite periodic crystals of vortices and has been used since the
early works of \cite{Abrikosov57, Kleiner64} where an ansatz for the
vortex lattice configuration was used to greatly restrict the degrees
of freedom contained in the $A_n$-parameters.  Infinite vortex
lattices in multi-component systems have been investigated with this
approach in several works including \cite{Kita02, Mueller02,
Reijnders04, Keifmmode06}.

The LLL approach is restricted to cases where the
magnetic length is small compared to the healing length: $\ell_B \ll
\xi$.  Away from this limit, the ground state wave functions will have
substantial components in higher Landau levels so the above projection
becomes unphysical.  On the other hand, most experiments on rotating
condensates of atomic gases have $\xi \ll \ell_B$.
This is the main restriction of the first method.

In the second approach, one considers $\omega_{\rm eff} >0$ so that the
condensate will have finite extent.  For this case one can discretise
and solve the imaginary-time GPE $\hbar \partial_\tau
\psi = - \delta E / \delta \psi^*$ where $\tau=it$ using numerical
techniques such as the the split-step Fourier method or the Crank-Nicolson
method (see \cite{Bao03}
and references therein).  In the long imaginary-time limit, such solutions will
generically approach the minimum of the Gross-Pitaevskii energy functional.  For
such calculations, one must choose grids sufficiently large so that the
wave function effectively vanishes at the boundary.  Due to this, the
boundary conditions involving the phase of $\psi$ are 
unimportant.  Direct numerical solution through this means has shown
several distinct vortex lattices structures for multi-component
systems as described in, for instance, the works of  \cite{Kasamatsu03, Barnett10, Mason11, Kuopanportti12}.

The main drawback of the aforementioned method when investigating
lattices of many vortices is that the underlying trap will obscure the
configuration of vortices \cite{Sheehy04}.  Therefore to infer the
ideal periodic configuration of vortices for the case of a spatially extended
condensate, one must require $\ell_B$ to be much smaller than the
condensate size.  In practice, this means one must numerically
investigate systems with at least $\sim100$ vortices.  Additionally,
since one must choose system sizes sufficiently large so that the wave
function effectively vanishes at the boundary, many computational grid
points are devoted to points of limited interest.

In this manuscript, we will describe an efficient numerical scheme to
determine ground state solutions of (\ref{energyfunctional}) which are
periodic in $|\psi|$.  We will focus on the case where $\omega_{\rm
eff}=0$ so that the condensate is spatially extended and an ideal
vortex lattice is expected to form.  This scheme does not have either
of the drawbacks of the two aforementioned methods for addressing
large vortex lattices.  In particular, motivated by the magnetic
translation group, we first introduce the so-called magnetic Fourier
transform and formulate the continuous GPE problem that yields $\psi$ with the correct periodic structure.  This correct problem arises since the magnetic Fourier transform naturally incorporates
twisted boundary conditions which must be satisfied for rotating
condensates (though they are unimportant for spatially localised
systems).  As a result, unnecessarily large computational domains can be avoided and systems sizes can be chosen on the order of the vortex lattice spacing.  We then introduce a discrete lattice model whose energy retains the same gauge symmetries as the continuous energy functional and converges to the energy functional as the lattice spacing decreases.  We show how spatial discretisation of this model can be achieved through a discrete version of the magnetic Fourier transform and how to compute it rapidly using standard fast Fourier transforms.  Further, we implement it with the well-known split-step time integration scheme routinely applied to
propagate the GPE in real and imaginary time \cite{Bao03}.  We test the method by showing that it
reduces to known results obtained in the lowest-Landau-level regime.
While we focus on the single-component case in this manuscript for
clarity and to make contact with known regimes, the framework can be
naturally extended to the multi-component case.

This work is organised as follows.  In Sec.\ \ref{Sec:2}, relevant background
on the magnetic translation group and gauge symmetries is given.  In
this Section the magnetic Fourier transform is introduced.  In Sec.\
\ref{Sec:3}, we describe how the spatial
components of (\ref{energyfunctional}) are discretised.  The
discretisation method allows one to access irrational aspect ratios of
the computational unit cell and leaves the gauge symmetry of 
(\ref{energyfunctional}) intact.
In Sec.\ \ref{Sec:4}, the
split-step
magnetic Fourier method is presented, which is the main result of the
present work.  Finally, in Sec.\ \ref{Sec:5}, applications of the method are
discussed and comparisons are made with known analytical results in
appropriate
regimes.

\section{Symmetries and Boundary Conditions for the rotated superfluid}
\label{Sec:2}

Consider the system of a rotating superfluid in the limit $\omega_{\rm
eff}=0$ (in the remainder of this work we will focus on this case).  In this limit, the superfluid will be spatially extended,
and a vortex lattice which is periodic in the density $\rho=|\psi|^2$
is formed.  To numerically address such a system, one would like to
choose a computational unit cell commensurate with that of the vortex
lattice containing a small number of vortices.  Though applying
periodic boundary conditions to the modulus of $\psi$ in this context
is clearly correct, the appropriate boundary conditions for its phase
are less apparent.  In the following, we will show that the latter can be deduced through use of the
so-called Magnetic Translation Operators (MTOs) \cite{Zak64_1,Zak64_2}.  
Since thorough expositions of the MTOs exist
elsewhere (see, for instance, \cite{Stone92}), we will only provide a
brief treatment of their properties relevant to our problem.  In the
following we will also describe the so-called Magnetic Fourier
Transform (MTF) which naturally incorporates the appropriate boundary
conditions of the problem, and will be utilised in following Sections.

\subsection{Background:  Gauge symmetry and Magnetic Translation Group}

We start with a few comments about the gauge symmetries of
(\ref{energyfunctional}).  This energy is invariant under the gauge
transformation $\psi \rightarrow e^{i \lambda/\hbar} \psi$, ${\bf
A}_s\rightarrow {\bf A}_s + \nabla \lambda$ where $\lambda$ is an
arbitrary function of the $xy$-coordinates.  Due to this invariance,
one is free to  choose a gauge which is best suited for the problem
at hand.  As defined in the previous section, ${\bf A}_s=\Omega m
(-y,x)$ is referred to as the symmetric gauge.  Another common choice
is the Landau gauge which is obtained by putting $\lambda=\Omega m xy$
so that ${\bf A}_l \equiv {\bf A}_s + \nabla \lambda= 2m\Omega (0,x)$.
In what follows we will write the vector potential in an arbitrary
gauge as ${\bf A}= {\bf A}_l + \nabla \lambda$, leaving $\lambda$
unspecified but noting that we may return to the original symmetric gauge
by putting $\lambda=-\Omega m xy$.  This will make apparent which
quantities are gauge invariant.  For instance, since the effective
magnetic field $B \equiv \epsilon_{ij} \partial_i A_j=2m\Omega$ (here
and after summation is implicit over repeated indices) is independent
of $\lambda$, it is a gauge invariant quantity.

The kinetic momentum operator entering (\ref{energyfunctional}) is
given by ${\bf P} \equiv {\bf p } - {\bf A}$ where ${\bf p}=-i \hbar
\nabla$ is the canonical momentum operator.  The generators of
magnetic translation, defined to be $\Pi_x=p_x-By-\partial_x \lambda$
and $\Pi_y=p_y-\partial_y \lambda$, commute with the kinetic momentum
operators: $[\Pi_i,P_j]=0$.  From these, one defines the magnetic
translation operators as $T({\bf R}) = e^{\frac{i}{\hbar}{\bf \Pi
\cdot R}}$.  When acting on quantities involving only coordinates, one
finds $T({\bf R}) f({\bf r}) T^{-1}({\bf R}) = f({\bf r} + {\bf R})$,
in complete analogy with the standard translation operators:
$\tilde{T}({\bf R}) \equiv e^{\frac{i}{\hbar}{\bf p \cdot R}}$.
However, unlike the standard translation operators, different MTOs do
not generally commute.  In particular, by noting that $[\Pi_x,
\Pi_y]=-i \hbar B$, one finds the commutator
\begin{align}
\label{commutator}
[T({\bf R}),T({\bf R'})] = 2i e^{\frac{i}{\hbar}{\bf \Pi}\cdot({\bf R}+{\bf R'})} \sin\left( \frac{B}{2 \hbar} \epsilon_{ij} R_i R_j' \right).
\end{align}

Solutions which minimise $(\ref{energyfunctional})$ will satisfy the
time-independent GPE: $\delta E / \delta
\psi^*=0$.  When $\omega_{\rm eff}=0$ this can be written as ${\cal H}
\psi = \mu \psi$ where $ {\cal H} = \frac{1}{2m} P^2+ g |\psi|^2.  $
For periodic vortex lattices, one has the symmetry $[{\cal H}, T({\bf
R})]=0$ when ${\bf R}$ is any lattice vector of the vortex lattice.
This can be used to greatly simplify the analysis.  However, note that
since different MTOs do not generally commute, the ground state $\psi$
will not generally be an eigenstate of all lattice MTOs.

\subsection{Twisted Boundary Conditions and Magnetic Fourier Transform}

We choose a rectangular computational unit cell of dimensions $L_x \times L_y$.  
Through full analogy with the treatment of a particle in a periodic potential in the absence of gauge fields, 
it is most natural to impose the following boundary conditions on $\psi$:
\begin{align}
\label{TBCx}
\psi(x,y) &= T(L_x {\bf \hat{x}}) \psi(x,y) = e^{-\frac{i}{\hbar}ByL_x}e^{-\frac{i}{\hbar}[\lambda(x+L_x,y)-\lambda(x,y)]}\psi(x+L_x,y) \\
\label{TBCy}
\psi(x,y) &= T(L_y {\bf \hat{y}}) \psi(x,y) = e^{-\frac{i}{\hbar}[\lambda(x,y+L_y)-\lambda(x,y)]}\psi(x,y+L_y).
\end{align}
These boundary conditions, to be contrasted with the more conventional periodic boundary conditions, 
reveal the acquisition of a phase `twist' over a period, and were found to be necessary
in early work \cite{Yang61}. For this reason we will refer to them as twisted boundary conditions (TBC) \cite{tHooft79}.

The above boundary conditions impose an important constraint on the
effective magnetic field $B$.  Specifically, by (\ref{TBCx}) and
(\ref{TBCy}) we must have $ \psi(x,y) =T(L_x {\bf \hat{x}}) T(L_y {\bf
\hat{y}}) \psi(x,y) = T(L_y {\bf \hat{y}}) T(L_x {\bf
\hat{x}})\psi(x,y) $ and so $[T(L_x {\bf \hat{x}}), T(L_y {\bf
\hat{y}})] \psi(x,y)=0$.  By comparison with (\ref{commutator}), one sees that
we have an imposed quantisation condition on the effective magnetic field: $B=\frac{2\pi \hbar }{L_x L_y} n$ where $n$ is
an integer.  Further insight can be gained by using
the Feynman relation \cite{Feynman55}, which expresses the density
of vortices as $\rho_v = \frac{m \Omega}{\pi \hbar}$. With this, the quantisation condition
becomes $\rho_v L_x L_y = n$.  Thus, we see that the integer
$n$ corresponds simply  to the number of vortices in the computational unit cell.

The conventional Fourier transform allows us to expand $\psi$ in a basis of
functions which are eigenstates of the canonical momentum operators
and which satisfy periodic boundary conditions.  Here we instead
expand in terms of eigenfunctions of the generators of magnetic
translation and require the TBCs to be satisfied.  This leads to the
following expressions defining the magnetic Fourier transform
and its inverse:
\begin{align}
\label{MFT}
\begin{array}{ll}
\tilde{\psi}(k_x,y) = \frac{1}{\sqrt{L_x}} \int_0^{L_x} dx \, e^{-i\left[k_x x+\frac{B}{\hbar} xy-\frac{1}{\hbar} \lambda(x,y)\right]}   \psi(x,y) \;\;\; & 
\tilde{\psi}(x,k_y) = \frac{1}{\sqrt{L_y}} \int_0^{L_y} dy e^{-i\left[k_y y-\frac{1}{\hbar} \lambda(x,y)\right]}   \psi(x,y)  \\
\psi(x,y) = \frac{1}{\sqrt{L_x}} \sum_{k_x}   e^{i\left[k_x x+\frac{B}{\hbar} xy -\frac{1}{\hbar} \lambda(x,y)\right]}  \tilde{\psi}(k_x,y) &
\psi(x,y) = \frac{1}{\sqrt{L_y}} \sum_{k_y}   e^{i\left[k_y y -\frac{1}{\hbar} \lambda(x,y)\right]}  \tilde{\psi}(x,k_y)
\end{array}
\end{align}
where $k_x$ and $k_y$ are summed over integer multiples of
$\frac{2\pi}{L_x} $ and $\frac{2\pi}{L_y}$ respectively.  We see that
these expressions have the desired properties.  In particular,
$
\Pi_x \tilde{\psi}(k_x,y) = k_x \tilde{\psi} (k_x,y),
$
$
\Pi_y \tilde{\psi}(x,k_y) = k_y \tilde{\psi} (k_x,y),
$
and the boundary conditions (\ref{TBCx}) and (\ref{TBCy}) are naturally satisfied.  

\section{Discrete Model}
\label{Sec:3}

Our aim is to minimise the energy of our system (\ref{energyfunctional})
subject to boundary conditions (\ref{TBCx}) and (\ref{TBCy}).
To do so we will numerically solve 
the imaginary-time GPE, $-\hbar \partial_\tau \psi = \delta E/\delta
\psi^*$ subject to TBCs.
Written out in full, the continuous equation is
\begin{align}
\label{TDGPE}
-\hbar \partial_\tau \psi =\frac{1}{2m} \left(-i\hbar\nabla-{ \bf A}_l - \nabla \lambda \right)^2 \psi + g |\psi|^2 \psi - \mu \psi
\equiv ({\cal H} -\mu)\psi.
\end{align}
In the long imaginary-time limit, $\psi$ will generically converge to the ground state wave function which minimises (\ref{energyfunctional}).
In particular one can show that the energy is a monotonically decreasing function of $\tau$
with
\begin{align}
\label{decay}
\hbar \frac{\text{d}E}{\text{d}\tau}=-2 \langle  ({\cal H}-\mu)^2 \rangle \le 0
\end{align}
where brackets denote the expectation value with respect to $\psi$. 
As can be seen from (\ref{decay}), the evolution will stop when $\psi$ becomes an eigenstate of ${\cal H}$ with eigenvalue $\mu$, i.e.\
a solution of the time-independent GPE.

We are thus faced with the problem of discretising the spatial components of
(\ref{TDGPE}) and note that there are multiple ways of doing so.  One approach is to treat (\ref{TDGPE}) by introducing discrete versions of the differential operators appearing in this equation.  While this approach is commonly adopted (see for example \cite{Bao03}), an alternative route and the one we use in this work, is to introduce an appropriate energy defined over a lattice and then, by taking its derivative with respect to the complex conjugate of the wave function values at the lattice points, obtain a discretised version of (\ref{TDGPE}).  By doing this, we can require that $(i)$ the discrete Hamiltonian matrix is Hermitian and $(ii)$
the discrete model inherits the exact gauge symmetry of the continuum
model.  The resulting model we find will be an extension of the
Hofstadter model \cite{Hofstadter76} which describes a particle on a square lattice under an effective
magnetic field.  That the resulting generalised model converges to the continuum theory (\ref{energyfunctional})
is perhaps not very surprising since it is well known that the long-wavelength theory
of the Hofstadter model is that of a continuum particle under the presence of a
perpendicular magnetic field.   The main technical innovation presented below is the use of
anisotropic tunnelling which enables one to treat computational unit cells of irrational
aspect ratios ${\cal R}=L_x/L_y$.  Accessing such irrational aspect ratios is necessary to properly
describe many important vortex configurations including the triangular lattice.

We introduce a grid of $N_x \times N_y$ points in a rectangular lattice
covering the computational unit
cell.  Points in this grid take on values ${\bf r} = a_x n {\bf \hat{x}} + a_y m {\bf \hat{y}}$
where $n$ and $m$ are integers satisfying $1\le n \le N_x$ and $1 \le
m \le N_y$ and $a_x= L_x/N_x$, 
$a_y=L_y/N_y$ give the spatial discretisation spacing (or lattice constant).
We next introduce the discrete wave function $\psi_{n,m}$ defined at each grid point and
where convenient we will use the alternative notation
$\psi_{\bf r}= \psi_{n,m}$.
We allow tunnelling of particles   between 
nearest-neighbour sites, and further impose an on-site repulsive interaction. 
Gauge fields are correctly incorporated into
lattice models  
 through the so-called Peierls substitution \cite{Peierls33, Hofstadter76}
where complex phases are incorporated into the tunnelling matrix elements.  
In particular to incorporate the gauge field into the tunnelling of particles
from site ${\bf r'}$ to site ${\bf r}$ we apply the replacement
$\psi_{\bf r}^* \psi_{\bf r'} \rightarrow  e^{\frac {i}{\hbar} \int_{\bf r'}^{\bf r} {\bf A} \cdot d{\bf r} } \psi_{\bf r}^* \psi_{\bf r'} $.    We note that, importantly, with this replacement the discrete model will
inherit the
exact gauge symmetry of the continuum model.

With the above considerations, we posit the following discrete expression for
the energy
\begin{align}
\notag
E_d =-\sum_{n,m} &\left[ w_1 e^{i(\lambda_{n,m} - \lambda_{n+1,m})} \psi_{n,m}^* \psi_{n+1,m} +
w_2 e^{i(\lambda_{n,m} - \lambda_{n,m+1})} e^{-i {\cal B}n} \psi_{n,m}^* \psi_{n,m+1} +{\rm c.c.} \right]
 \\
 \label{discreteE}
&+\sum_{n,m}\left[ \frac{U}{2} |\psi_{n,m}|^4 - \tilde{\mu} |\psi_{n,m}|^2 \right]
\end{align}
where $w_1$, $w_2$, and $U$  are real positive parameters having units of energy, ${\cal B}$ is the lattice magnetic field giving the net flux
per lattice plaquette, $\tilde{\mu}$ is a chemical potential for the discrete model, and the 
real parameters $\lambda_{n,m}$
reflect the gauge freedom of the problem.  
When $U=\lambda_{n,m}=0$ and $w_1=w_2$,  the discrete model reduces to  the
well-known Hofstadter model \cite{Hofstadter76}.  In this sense, the above expression (\ref{discreteE})
can be viewed as a generalised Hofstadter model.

To make contact with the continuum theory (\ref{energyfunctional}), 
we relate the discrete and continuum wave functions as
$
\psi_{n,m} = \sqrt{a_x a_y} \psi(a_x n,a_y m).
$
With this, one can verify that the discrete energy reduces to the
continuum energy 
provided we make the following identifications: 
$w_1 a_x^2=w_2 a_y^2 = \frac{\hbar^2}{2m}$, $U=g / a_x a_y$, ${\cal B} = B a_x a_y / \hbar$, 
$\lambda_{n,m} = \lambda(a_x n,a_y m)/\hbar$,
and
$\tilde{\mu} = \mu - 2w_1-2w_2$.   
The resulting discretisation error is found to be second order: $E- E_d=\mathcal{O}(a_x^2) + \mathcal{O}(a_y^2)$.
Importantly, we note that the aspect ratio of the computational unit cell is  given by
${\cal R} = \frac{L_x}{L_y} = \frac{N_x}{N_y} \sqrt{\frac{w_2}{w_1}}.$  Therefore, with anisotropic hopping we may access
computational unit cells with irrational aspect ratios which are necessary for addressing the triangular
vortex lattice, for instance.    
In order for the discrete theory to accurately 
describe the continuum theory, the discretisation lattice constants,
$a_x$ and $a_y$, must be the smallest length scales in the problem.
Specifically, in what follows we require $a_x, a_y \ll \xi, \ell_B$.    
On the other hand, we note that
 the discrete theory remains well-defined and physically relevant away from this limit.  
Indeed, most investigations of the Hofstadter model are focused on regimes where $a_{x}$ and $a_y$ are on the order of the magnetic 
length.  

The continuum expressions from the previous Section carry over
naturally to the discretised case, which we tabulate 
in the remainder of the present Section.  The discrete imaginary-time GPE is given by 
\begin{align}
\hbar \partial_\tau \psi_{n,m} = -\frac{\partial E_{d}}{\partial \psi^*_{n,m}}.
\end{align}
The discrete MFT, c.f.\ (\ref{MFT}), is
\begin{align}
\begin{array}{ll}
\tilde{\psi}_{k_x,m}= \frac{1}{\sqrt{N_x}} \sum_{n}  e^{-i\left[k_x n+ {\cal B} nm - \lambda_{n,m} \right]}   \psi_{n,m} \;\;\;\;\; & 
\tilde{\psi}_{n,k_y} = \frac{1}{\sqrt{N_y}}  \sum_{m}  e^{-i\left[k_y m -  \lambda_{n,m}\right]}   \psi_{n,m}  \\
\psi_{n,m} = \frac{1}{\sqrt{N_x}} \sum_{k_x}   e^{i\left[k_x n+{\cal B} nm - \lambda_{n,m} \right]}  \tilde{\psi}_{k_x,m} &
\psi_{n,m} = \frac{1}{\sqrt{N_y}} \sum_{k_y}   e^{i\left[k_y m -  \lambda_{n,m}\right]}  \tilde{\psi}_{n,k_y}.
\end{array}
\label{MFTd}
\end{align}
Here, $n$ and $m$ take on integer values for which $1 \le n \le N_x$ and $1 \le m \le N_y$ respectively.
The wave numbers
$k_x$ and $k_y$ take on integer multiple values of $\frac{2\pi}{N_x}$ and $\frac{2\pi}{N_y}$ 
for which $\frac{2\pi}{N_x} \le k_x \le 2\pi$ and $\frac{2\pi}{N_y} \le k_y \le 2\pi$ respectively.
Finally, the twisted boundary conditions, c.f.\  (\ref{TBCx}) and (\ref{TBCy}), in discretised form
are
\begin{align}
\psi_{n+N_x,m} &=   e^{i {\cal B} m N_x}e^{i [\lambda_{n+N_x,m}-\lambda_{n,m}]}\psi_{n,m} \\
\psi_{n,m+N_y} &=e^{i [\lambda_{n,m+N_y}-\lambda_{n,m}]}\psi_{n,m}.
\end{align}

\section{The Split-Step Magnetic Fourier Method}
\label{Sec:4}
Split-step spectral methods provide a versatile means to solve a number of linear and 
non-linear differential equations \cite{Taha84}.  As a simplest case,
consider a (linear  and real-time) Schrodinger equation with Hamiltonian composed of two non-commuting operators
$A$ and $B$ which each have known spectra: 
$i \hbar \partial_t \psi = (A+B) \psi$.
The wave function is advanced by a time step through $\psi(t+\Delta t) = e^{-\frac{i}{\hbar}(A+B)\Delta t} \psi(t)$.
For cases where the Hamiltonian operator $A+B$ is difficult to diagonalise, 
it is beneficial to employ the Strang splitting \cite{Strang68} to propagate in time:
\begin{align}
\label{propagate}
\psi(t+\Delta t) =  e^{-\frac{i}{\hbar}(A+B)\Delta t} \psi(t) = 
e^{-\frac{i}{\hbar}B\Delta t/2}e^{-\frac{i}{\hbar}A\Delta t}e^{-\frac{i}{\hbar}B\Delta t/2}\psi(t) +\mathcal{O}\left(\Delta t^3\right).
\end{align}
One can then efficiently perform the operation on the RHS of 
(\ref{propagate})
by successively transforming $\psi$ between the eigenbasis of $A$ and $B$ and performing Hadamard products.
Most often $A$ is diagonal in the momentum basis while $B$ is diagonal in the position basis (or vice-versa) so the
efficiency of the spit-step method relies on that of the fast Fourier transform algorithm.  Indeed, the split-step Fourier method is now a standard approach for
numerically solving the GPE (see \cite{Bao03}
and references therein).  We show here how this powerful
method can also be applied to our problem by instead using the magnetic Fourier transform to capture the twisted
boundary conditions given in (\ref{TBCx}) and (\ref{TBCy}).

With the above in mind, we wish to write the energy $E_d$ as a sum of diagonalised components.  As stated, this can be achieved
with the machinery of the magnetic Fourier transform introduced in the previous sections.   In particular, using 
(\ref{MFTd}) we can write $E_d = E_{x} + E_{y}+E_{{\rm int}}$ where
\begin{align}
\label{Ex}
E_x &= 4 w_x \sum_{k_x,m} \sin^2\left( \frac{k_x + {\cal B} m}{2}  \right) |\tilde{\psi}_{k_x,m}|^2 \\
\label{Ey}
E_y &= 4 w_y \sum_{n,k_y} \sin^2\left( \frac{k_y-{\cal B}n }{2} \right) |\tilde{\psi}_{n,k_y}|^2 \\
\label{Eint}
E_{\rm int} &= \sum_{n,m} \left(  \frac{U}{2} |\psi_{n,m}|^4 - \mu |\psi_{n,m}|^2 \right).
\end{align}
Note that the gauge fields $\lambda_{n,m}$ are absent in the above expressions, making the
gauge invariance of each quantity apparent.  

Because the above energy is composed of three  pieces, the Strang splitting needs 
to be applied twice.  
The split-step 
MFT procedure, to advance the wave function $\psi(\tau)$  evolving through the imaginary-time 
GPE by a single time step $\Delta \tau$,
then proceeds with the following computations:
\begin{enumerate}
	\item $\psi_1=e^{-H_{\text{int}} (\rho_a)
            \frac{\Delta \tau}{2\hbar}}\psi(\tau)$
	\item $\psi_2=\mathcal{MFT}_x^{-1}\left[e^{-H_x \frac{\Delta \tau}{2\hbar}}\mathcal{MFT}_x\left[\psi_1\right]\right]$
	\item $\psi_3=\mathcal{MFT}_y^{-1}\left[e^{-H_y  \frac{\Delta\tau}{\hbar}}\mathcal{MFT}_y\left[\psi_2\right]\right]$
	\item $\psi_4=\mathcal{MFT}_x^{-1}\left[e^{- H_x \frac{\Delta\tau}{2\hbar}}\mathcal{MFT}_x\left[\psi_3\right]\right]$
	\item $\psi(\tau+\Delta \tau )=e^{- H_\text{int}(\rho_b)  \frac{\Delta\tau}{2\hbar}}\psi_4$
\end{enumerate}
where $\mathcal{MFT}_x$ and $\mathcal{MFT}_y$ are the operations defined in \eqref{MFTd}. 
The quantities in the exponents directly follow from (\ref{Ex}),
(\ref{Ey}), and ({\ref{Eint}).  In particular, 
$H_x= 4w_x \sin^2\left( \frac{k_x + {\cal B} m}{2}  \right)$,
$H_y=4w_y \sin^2\left( \frac{k_y-{\cal B}n }{2} \right)$,
and $H_{\rm int} (\rho) = U \rho - \mu$.  
Thusfar we have not specified the densities $\rho_a$ and $\rho_b$
entering $H_{\rm int}$ in the steps above.  For certain choices
(e.g. putting $\rho_a=\rho_b=|\psi(\tau)|^2$), the method will lose its
second-order accuracy.  To determine values of   $\rho_a$ and
$\rho_b$ for which the method will retain its second-order accuracy
(as in the linear case), one can compare $\psi(\tau+\Delta \tau)$
computed with the above algorithm with the second-order Taylor
expansion 
$
\psi(\tau + \Delta \tau) \approx \psi(\tau) + \partial_\tau \psi(\tau)
\Delta \tau + \frac{1}{2} \partial^2_\tau \psi(\tau) (\Delta \tau)^2
$
where the imaginary-time GPE can be used to evaluate the coefficients in
this expansion.  
Through this comparison, one finds that with the choices 
\begin{align}
\label{densities}
\rho_a = |e^{-H_{\rm int}(|\psi(\tau)|^2) \frac{\Delta
    \tau}{2\hbar}} \psi(\tau)|^2, \;\; \rho_b = |\psi_4|^2,
\end{align}
the method will be second-order accurate.
The determination of the appropriate densities entering the algorithm
is akin to the methods presented  in \cite{Javanainen06}.  There, it
was determined that, provided the most updated densities were used,
the split-step method for the real-time GPE would be second-order
accurate.  We note, however, that this result does not carry over to
the imaginary-time GPE in that if one chooses $\rho_a = |
\psi(\tau)|^2$ and $\rho_b = |\psi_4|^2$, the method will only be
first-order 
accurate.

It is often desirable to propagate the imaginary-time GPE for the case
where the particle-number normalisation $\int |\psi|^2 d x\, dy=\int
 \rho\, dx \, dy={\cal N}$ 
is fixed for all times during the evolution.  This constraint can be achieved
through the use of a time-dependent chemical potential entering the
imaginary-time
GPE.   The above algorithm can be naturally adapted to this scenario by applying
normalisations at appropriate places in the above steps.  In
particular, after computing the densities $\rho_a$ and $\rho_b$ 
using the expressions in (\ref{densities}), 
they should each be normalised to the total particle number ${\cal N}$.
Additionally, at the end of each time-step propagation, the resulting
wavefunction should be normalised.  Due to these normalisations, the
 chemical potential entering the algorithm becomes
a free parameter and can be set to zero.
The second-order accuracy of this fixed-${\cal N}$ 
method has been verified with numerical tests.

Note that the
 extension of the  split-step method from linear to non-linear equations, as done above,
 is of small influence on the computational time.   For instance, if we put $g=0$
making the GPE linear, the computational cost of the split-step method at
 leading order will not be affected. 
 The MFT can be  implemented in a straightforward way by using existing fast Fourier
 transform packages
 as it can be written in terms of direct multiplications and Fourier transforms as:
\begin{align}
\mathcal{MFT}_x\left[\psi\right]=&\mathcal{F}_x\left[e^{-inm\mathcal{B}}e^{-i\lambda_{n,m}}\psi\right]\\
\mathcal{MFT}_y\left[\psi\right]=&\mathcal{F}_y\left[e^{-i\lambda_{n,m}}\psi\right]
\end{align}
where $\mathcal{F}_x$ and $\mathcal{F}_y$ denote the standard Fourier transforms.
Thus  the MFT  algorithm is as fast as the conventional fast Fourier transform to leading order.
More specifically, for $N$ discretisation points, the method has $N\log N$ computational cost.
Finally, we note that the above scheme can be used to simulate the real-time GPE with a Wick rotation where one puts $\tau \rightarrow it$.

\section{Numerical Tests and Discussion}
\label{Sec:5}

In this Section, we will provide some preliminary applications of the split-step magnetic Fourier
method, showing how it can reproduce known results in appropriate regimes and also how these results can be extended. 
We start by considering the lowest lowest-Landau-level regime, since this case has
several known results with which we can compare.

To characterise vortex lattices, following \cite{Abrikosov57}, it is helpful to introduce the
dimensionless  parameter 
\begin{align}
\label{beta}
\beta=A\frac{\int|\psi|^4dxdy}{\left(\int|\psi|^2dxdy\right)^2}
\end{align}
where $A=L_xL_y$ is the area of the computational unit cell. This parameter is of particular interest because it can be directly related to the interaction energy of the system as $\frac{g}{2}\beta\frac{\mathcal{N}^2}{A}$, with $\mathcal{N}=\int|\psi|^2dxdy$.
In the LLL regime the remaining terms in the energy are quenched, thus minimising the energy
is equivalent to minimising $\beta$.
When the number of vortices is restricted to two per computational unit cell, one can compute $\beta$ analytically as a function of the aspect ratio $\mathcal{R}$ \cite{Kleiner64}
\bes{
\beta^A(\mathcal{R})=\sqrt{\frac{\mathcal{R}}{2}}\left(f_0^2+2f_0f_1-f_1^2\right),
}{betaR}
with
\bes{
f_n=\sum_{m=-\infty}^\infty e^{-\pi \mathcal{R}(2m+n)^2/2}
}{fns}
where the vortices are placed within the unit cell so as to maximise the separation
between neighbouring vortices.  Recall that a unit cell commensurate with that of
the ground state vortex lattice of infinite spatial extent must be chosen to obtain
the ground state energy.  Unit cell sizes differing from this will introduce frustration.
Therefore $\beta$ should be minimised with respect to ${\cal R}$.
One expects minima to occur at  aspect ratios
which are commensurate with a triangular vortex lattice \cite{Kleiner64, Tkachenko66}.  Since $\beta(\mathcal{R})=\beta(\mathcal{R}^{-1})$
we will restrict our attention to $\mathcal{R}\ge 1$.

We now turn to numerically computing $\beta$ using the split-step magnetic Fourier method.  
For fixed values of $\mathcal{R}$ and $\ell_B/\xi$, 
starting with an
initial randomised state, the imaginary-time GPE is evolved on a $256 \times 256$ grid
until a time-independent state is obtained.  Convergence as a function of the time
step $\Delta \tau$ is also checked.
\begin{figure}[h!]
	\centering
	\newlength\figureheight 
	\newlength\figurewidth 
	\setlength\figureheight{6cm} 
	\setlength\figurewidth{10cm}
	\includegraphics{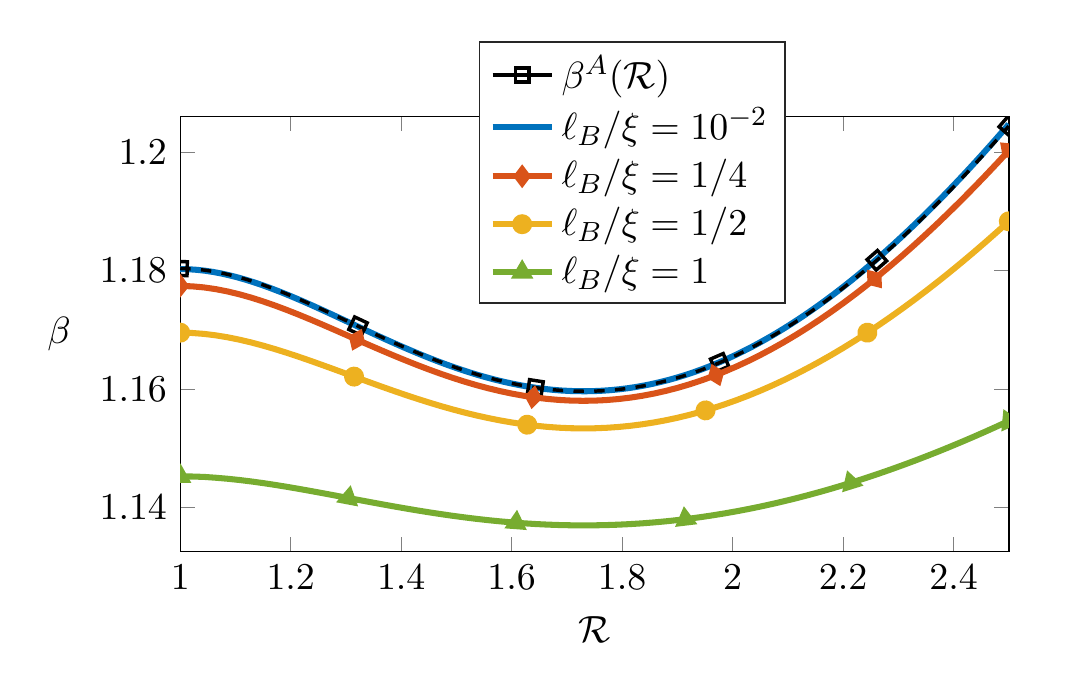};
	\caption{Numerical results of the dimensionless parameter $\beta$ (\ref{beta}) characterising the interaction energy of the system  as a function of the aspect ratio of the computational unit cell.  Excellent agreement is found between the numerical results and the analytical expression $\beta^A$ in the lowest Landau level regime.}
	\label{betanums}
\end{figure}
In Fig.~\ref{betanums}, several curves of $\beta$ versus the aspect ratio $\mathcal{R}$
are shown for different values of $\ell_B/\xi$ for systems containing two vortices per computational unit cell.  
In the limit $\ell_B \ll \xi$ one finds excellent agreement 
with the LLL analytical expression $\beta^A$ as expected.   The scheme naturally allows
one to extend beyond the LLL regime for which simple analytical expressions for $\beta$
are not available.  As $\ell_B/\xi$ is increased, one finds that $\beta$ decreases, reflecting
the system's tendency towards a nearly uniform density (apart from the vortex cores) in
the large interaction limit.
Also, as expected, the minimum for all
curves occurs at ${\cal R} = \sqrt{3}$, which is commensurate with the triangular vortex lattice.  A local maximum occurs at ${\cal R} = 1$ which corresponds to the square vortex lattice.  

Because the energy of the system is an extensive quantity, one might expect the
energy per particle, $\tilde{E} \equiv \frac{E+\mu\mathcal{N}}{\mathcal{N}}$, to be unchanged when 
the size of the computational cell  is increased ceteris paribus (for convenience we have removed the term
involving the chemical potential in (\ref{energyfunctional})).
This is only true, however, for computational unit cells commensurate with the ground state
vortex configuration as
increasing the size of the cell while keeping the vortex density fixed
allows the system to have more  degrees of freedom.  In the following, we consider doubling the spatial dimensions of the computational
unit cell from one containing two vortices to one containing eight vortices.  From general principles
one will have ${\tilde E}_8 \le \tilde{E}_2$ where $\tilde{E}_8$ and $\tilde{E}_2$ are the 
energies per particle of the systems with smaller and larger computational unit cells respectively.
In Fig.~\ref{extensive} the energy per particle obtained using the split-step magnetic Fourier method is shown for systems having two and eight vortices per computational unit cell. 
\begin{figure}[h!]
	\centering
	\setlength\figureheight{6cm} 
	\setlength\figurewidth{10cm}
	\includegraphics{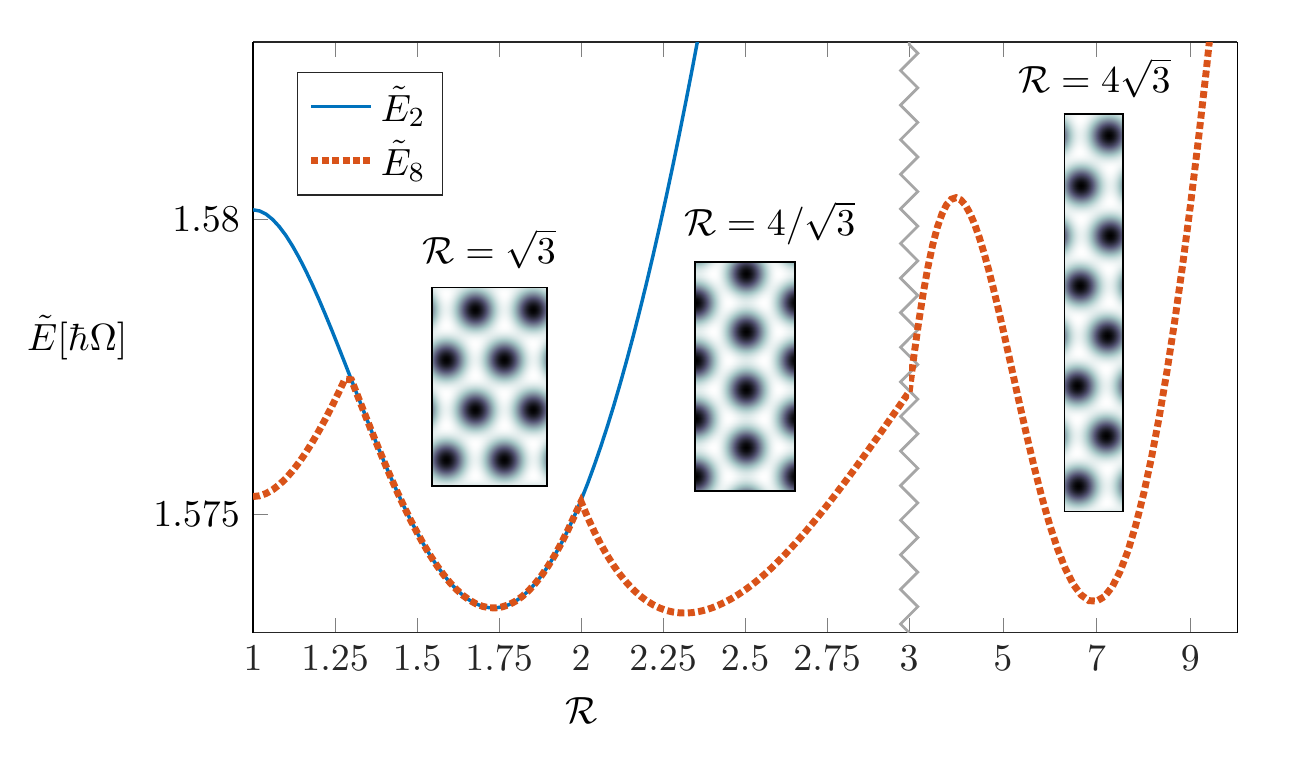};
	\caption{The energy per particle of systems with two and eight vortices in the computational unit cell, $\tilde{E}_2$ and $\tilde{E}_8$, as a function of the ratio $\mathcal{R}$. The former provides an upper  bound for the latter. The triangular lattices corresponding to the ground states are degenerate and the arrangements of the vortices in each of the ground states of the eight-vortex system are shown.}
	\label{extensive}
\end{figure}
As expected, we have
${\tilde E}_8 \le \tilde{E}_2$ for all curves.  The curves coincide
near $\mathcal{R}=\sqrt{3}$ which is commensurate with the triangular
vortex lattice 
in both cases.  The system with eight vortices can
also achieve a triangular vortex lattice at aspect ratios $\mathcal{R}=4/\sqrt{3},4\sqrt{3}$ and we note that degenerate minima of $\tilde{E}_8$ occur at these values.
The computation has been carried out for $\ell_{{B}}/\xi=1$ but the same results--namely $\tilde{E}_2$ being an upper bound for $\tilde{E}_8$, their equivalence close to $\mathcal{R}=\sqrt{3}$ and the existence of the three degenerate minima for $\tilde{E}_8$--hold for any ratio $\ell_{{B}}/\xi$.

\section{Outlook and conclusions}
\label{Sec:6}
In conclusion, we have described a computational method which can efficiently find the minimum energy of an
infinite vortex lattice within Gross-Pitaevskii mean field theory.  We have shown how to extend the conventional
split-step Fourier method to include twisted boundary conditions
through use of the magnetic translation group.  We have
tested the method for particular cases, and showed that it reproduces known results in the lowest Landau level regime.
The method can be extended naturally to treat multi-component systems such as condensate mixtures and spinor condensates, as
well as multicomponent condensates under more general synthetic gauge fields.  
Such systems can have richer vortex lattice structures, and applying the methodology described in this work to them
warrants further investigation.

\begin{acknowledgements}
This work was supported in part by the European Union's Seventh Framework Programme for
research, technological development, and demonstration
under Grant No.\ PCIG-GA-2013-631002.
\end{acknowledgements}

\bibliographystyle{apsrev4-1}

%

\end{document}